\documentclass{aa}
\usepackage{amssymb,amsfonts,amsmath,times,pict2e}
\usepackage{natbib}
\usepackage{graphicx}
\usepackage{enumerate}
\usepackage{subfigure}
\usepackage{bm}
\usepackage{hyperref}

\usepackage{color}
\usepackage{ulem}
\definecolor{mygray}{gray}{0.6}
\definecolor{magenta}{rgb}{0.858, 0.188, 0.478}
\usepackage{xspace}

\newcommand{\fg}[1]{Fig.~\ref{fig:#1}}
\newcommand{\Fg}[1]{Figure~\ref{fig:#1}}%beginning of the sentence
\newcommand{\eq}[1]{Eq.~(\ref{eq:#1})}
\newcommand{\Tb}[1]{Table~\ref{tab:#1}}
\newcommand{\Eq}[1]{Equation~(\ref{eq:#1})}%beginning of the sentence
\newcommand{\Eqs}[2]{Equations~(\ref{eq:#1})--(\ref{eq:#2})}%beginning of the sentence
\newcommand{\eqs}[2]{Eqs.~(\ref{eq:#1})--(\ref{eq:#2})}%beginning of the sentence
\newcommand{\app}[1]{App.~\ref{sec:#1}}

\newcommand{\se}[1]{Sect.~\ref{sec:#1}}
\newcommand{\tb}[1]{Table~\ref{tab:#1}}%beginning of the sentence

\newcommand{\micr}{$\mu$m\xspace}

\newcommand{\mgsio}{MgSiO$_3$\xspace}

\begin{document}

\title{ARCiS framework for exoplanet atmospheres}
\subtitle{cloud transport model}
\titlerunning{Exoplanet atmosphere cloud transport model}

\author{Chris W. Ormel \inst{1} and Michiel Min\inst{2} }

\institute{Anton Pannekoek Institute (API), University of Amsterdam, Science Park 904,1090 GE Amsterdam, The Netherlands
\label{inst1}
\and
Netherlands Institute for Space Research (SRON), Sorbonnelaan 2, 3584 CA Utrecht, The Netherlands
\label{inst2}
\\
\email{[c.w.ormel@uva.nl,m.min@sron]}
}

\date{\today}

\abstract{Understanding of clouds is instrumental in interpreting current and future spectroscopic observations of exoplanets. modeling clouds consistently is complex, since it involves many facets of chemistry, nucleation theory, condensation physics, coagulation, and particle transport.}
{We aim to develop a simple physical model for cloud formation and transport, efficient and versatile enough that it can be used, in modular fashion for parameter optimization searches of exoplanet atmosphere spectra. In this work we present the cloud model and investigate the dependence of key parameters as the cloud diffusivity $K$ and the nuclei injection rate $\dot\Sigma_n$ on the planet's observational characteristics.}
{The transport equations are formulated in 1D, accounting for sedimentation and diffusion. The grain size is obtained through a moment method. For simplicity, only one cloud species is considered and the nucleation rate is parametrized.
From the resulting physical profiles we simulate transmission spectra covering the visual to mid-IR wavelength range.}
{We apply our models toward KCl clouds in the atmosphere of GJ1214 b and toward \mgsio clouds of a canonical hot-Jupiter. We find that larger $K$ increases the thickness of the cloud, pushing the $\tau=1$ surface to a lower pressure layer higher in the atmosphere. A larger nucleation rate also increases the cloud thickness while it suppresses the grain size. Coagulation is most important at high $\dot\Sigma_n$ and low $K$. We find that the investigated combinations of $K$ and $\dot\Sigma_n$ greatly affect the transmission spectra in terms of the slope at near-IR wavelength (a proxy for grain size), the molecular features seen at $\sim$$\mu$m (which disappear for thick clouds, high in the atmosphere), and the $10\,\mu\mathrm{m}$ silicate feature, which becomes prominent for small grains high in the atmosphere.}
{Clouds have a major impact on the atmospheric characteristics of hot-Jupiters, and models as those presented here are necessary to reveal the underlying properties of exoplanet atmospheres. The result of our hybrid approach -- aimed to provide a good balance between physical consistency and computational efficiency -- is ideal toward interpreting (future) spectroscopic observations of exoplanets.}

\keywords{Planets and satellites: atmospheres --- Planets and satellites: composition --- Methods: numerical}

\maketitle

\section{Introduction}
\noindent
The composition of exoplanet atmospheres contains very important clues to their formation and evolution. Different formation scenarios predict different abundances of key elements like C, O, N, and Si \citep[e.g.,][]{2011ApJ...743L..16O,HellingEtal2014,MordasiniEtal2016,2017MNRAS.469.4102M}. Measuring the abundances of these elements is one of the major goals of performing exoplanet atmosphere spectroscopy \citep[see e.g.,][]{2017AJ....153...83B}. With the launch of the James Webb Space Telescope (JWST) scheduled in 2021, a new wavelength window, the near- to mid-IR, will open up for compositional analysis of exoplanet atmospheres. With the recently selected ARIEL mission on the 2028 horizon, performing spectroscopy of a statistically significant sample of exoplanets, the future for atmosphere characterization looks particularly bright \citep{2018ExA....46...45T}. This new spectroscopic window presents us with many opportunities, but at the same time provides challenges in proper interpretation.

One of the major hurdles in atmospheric characterization is the presence of clouds obscuring the gaseous content of the atmosphere. Besides shielding the gaseous atmosphere from detection, clouds also alter the chemical composition of the gaseous atmosphere. By removing elements from the gas phase and raining them down to deeper layers, cloud processes alter the chemical composition of the atmosphere.  For the interpretation of the atmosphere spectrum, this can lead to an incorrect assessment of the atomic composition of the bulk planet.

The difficulty of modeling cloud formation has led to a rich variety of different treatments of clouds. For models that retrieve key atmospheric parameters (temperature, pressure, and chemical profiles) directly from the observations, so-called retrieval models, it is very important that the simulations can be performed in the most computationally efficient manner. These methods often simply apply an atmospheric pressure below which the opacity of the atmosphere is gray (or infinite) with the possible addition of Rayleigh scattering haze \citep[see e.g.,][]{2015ApJ...814...66K, 2017ApJ...834...50B}. This assumption might be acceptable for the narrow wavelength range considered in most studies right now. However, when the wavelength range extends, it becomes crucial to take into account the wavelength dependence of the optical properties of the cloud particles. 

In forward models the complexity of the cloud formation varies. The approximate cloud formation model by \citet{AckermanMarley2001} is probably one of the most widely used cloud formation frameworks.  In this model the physical properties of the cloud particles are parameterized in terms of a single parameter, $f_\mathrm{sed}$, the ratio between the particle sedimentation and the turbulent eddy velocities. It can be regarded as a proxy of the cloud particle size, although for constant $f_\mathrm{sed}$ the size will vary with height. While the assumption of a constant $f_\mathrm{sed}$ is not a priori evident, the advantage of this approach is that it avoids an elaborate grain microphysical prescription. At the other extreme are full self-consistent models that follow the microphysics of grain nucleation, condensation, transport and chemistry \citep{HellingEtal2008,GaoEtal2018}.
Nevertheless, enhanced model complexity also introduces drawbacks. First, these models tend to be computationally demanding and are therefore not well suited for implementation in retrieval codes. In addition, increased model complexity often implies a great number of free parameters, which either need to be justified or else need to be explored, increasing the computational demand. 
Most crucially in this regard is the formation of condensation seeds (nucleation), which under the extreme conditions in exoplanet atmospheres is poorly understood. These considerations might argue in favor of building a retrieval framework that contains no cloud formation physics and, by fitting the spectrum, have the observations tell us what is going on \citep[see e.g.,][]{2017ApJ...834...50B, 2018AJ....155..156T}. While this is a widely-used approach, a drawback of this approach is that it comes with a plethora of free parameters, which physical consistency is not a priori guaranteed (e.g., the feedback of cloud formation on the atmospheric composition is not necessarily accounted for).

Here, we aim for an intermediate approach, in which the cloud structure is computed in a simplified but consistent forward model. We envision that such a hybrid model has the benefits of both worlds: it should include the most elementary cloud physics (e.g., condensation and cloud transport) consistently, but yet be be computationally fast and flexible enough to allow for parameter studies and incorporation in retrieval algorithms. Recent examples of this approach are the semi-analytical model by \citet{CharnayEtal2018}, applicable for Brown Dwarfs and young exoplanets, 1D dust coagulation models of atmospheres of planets embedded in their natal gas disk \citep{MovshovitzEtal2010,Mordasini2014,Ormel2014}, and 1D cloud transport models for exoplanets \citep{OhnoOkuzumi2017,OhnoOkuzumi2018,KawashimaIkoma2018}. A common characteristics of these approaches is that they are one dimensional and consider a single, representative particle size that varies with height. In this paper we follow these leads to efficiently compute the formation of clouds for hot Jupiters. We use a diffusion/condensation framework to compute the growth of cloud particles and include particle coagulation. On the other hand the nucleation rate is parameterized to accommodate the large uncertainty in nucleation efficiency. 

The cloud model that we present in this paper will become part of a general framework for analysis and retrieval of exoplanet spectra\footnote{The cloud model, written in \texttt{python}, is publicly available at \url{http://www.exoclouds.com/}.}. In this context we are developing a code for computation of atmospheric properties, radiative transfer, and retrieval named \texttt{ARCiS} (ARtful modeling Code for exoplanet Science). The overarching aim of \texttt{ARCiS} is to develop an approach that is well-balanced between physical consistency, model complexity and computationally efficiency. The physical consistency allows for direct physical interpretation of observations. The modest model complexity allows for in depth understanding of the effects going on. The computational efficiency ensures that the model can be efficiently used in spectral retrieval analysis of observations. In this paper we focus on the cloud model; a validation of the entire \texttt{ARCiS} framework and subsequent fitting of real spectra will be deferred to upcoming studies.

In \se{model} the cloud formation model is explained. In \se{results} we present the resulting cloud structures and transmission spectra for a sub-Neptune (GJ1214 b) and for a typical hot-Jupiter planet, while varying the diffusivity and nucleation rate. In \se{spectra}, we present the synthetic transmission spectra in the near- to mid-IR for the hot-Jupiter configuration. An assessment of the cloud model is proved in \se{assess}. In \se{summary} we summarize the results and discuss extensions to this modeling framework.
\section{Model}
Our cloud particle model entails solving for the 1D steady state solutions to the transport equations involving vapor, condensates, and nuclei.  Cloud particles are initiated through nucleation at prescribed rates. Vapor can condense on these seeds and the particles may further growth by coagulation. Particles are transported by gravitational settling and turbulent (eddy) diffusion, until they reach the bottom of the cloud, hot enough to result in their evaporation.  For simplicity a single species -- KCl in case of GJ1214 b and \mgsio in case of the generic hot Jupiter -- is considered. The choice for the species in question is arbitrary, although for the cloud to be observed, it must lie high in the atmosphere. Hence, the temperature of the upper atmosphere must be lower -- but not much lower -- than the condensation temperature. It is also straightforward to extend the model to include other chemical compounds.
\label{sec:model}
\begin{figure*}
    \centering
    \includegraphics[width=88mm]{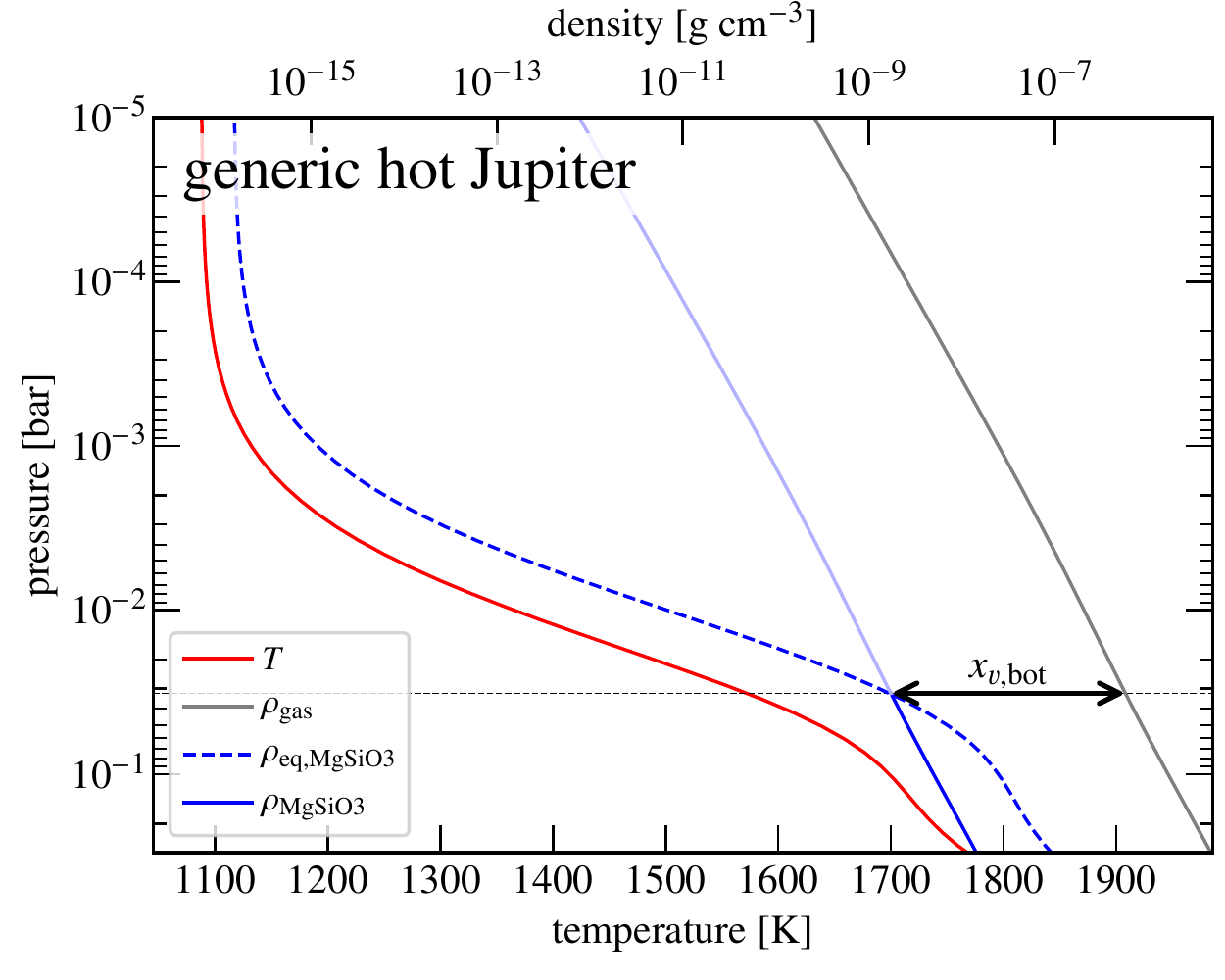}
    \includegraphics[width=88mm]{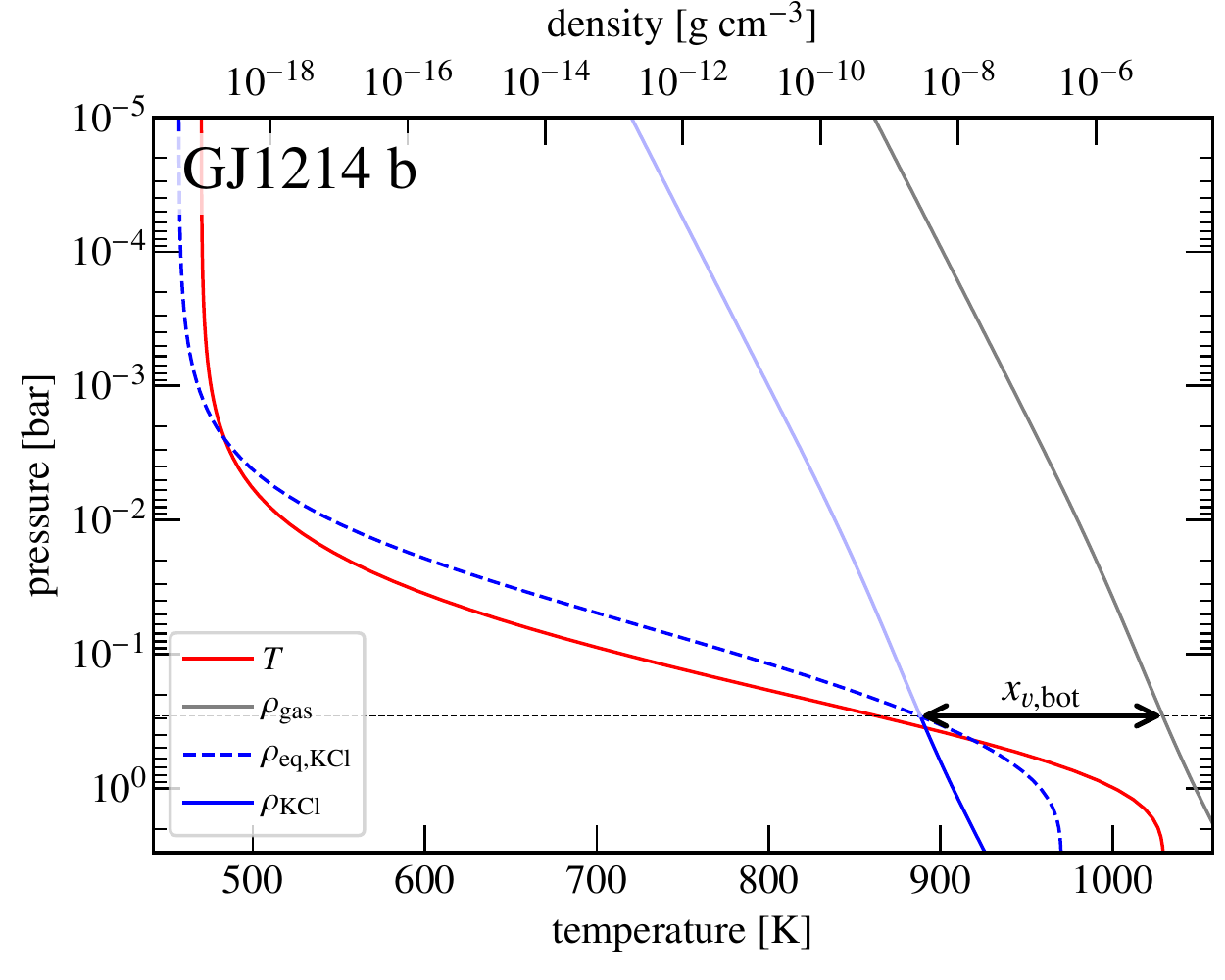}
    \caption{Left: the physical structure of a generic hot-Jupiter atmosphere. The temperature $T(P)$ and gas density profile $\rho_\mathrm{gas}(P)$ are obtained from the \citet{Guillot2010} atmosphere model. The equilibrium density $\rho_\mathrm{eq}$ of \mgsio is obtained from \eq{Psat}. Below the cloud deck (horizontal dashed line), $\rho_v=\rho($\mgsio$)<\rho_\mathrm{eq}$; the species are present at constant abundance, such that $x_v = \rho_v/\rho_\mathrm{gas} \equiv x_\mathrm{v,bot}$ is constant. Above the cloud base ($\rho_v>\rho_\mathrm{eq}$; light blue) cloud formation (not included in this figure) will reduce the vapor density at the expense of condensates. Right: same for KCl in the GJ1214 b atmosphere.}
    \label{fig:HJ-profile}
\end{figure*}

\subsection{Atmosphere model}
We consider the atmosphere typical to a hot Jupiter planet.  To obtain its physical structure -- the temperature $T(z)$, pressure $P(z)$ and density $\rho(z)$ profiles -- we adopt the atmosphere model of \citet{Guillot2010} to obtain a relation between temperature and depth:
\begin{multline}
    \label{eq:Guillot}
    T^4 = \frac{3T_\mathrm{int}^4}{4} \left( \frac{2}{3}+\tau \right) \\ +\frac{3T^4_\mathrm{irr}f_\mathrm{irr}}{4}\left[ \frac{2}{3} +\frac{1}{\gamma\sqrt{3}} +\left( \frac{\gamma}{\sqrt{3}} -\frac{1}{\gamma\sqrt{3}} \right) e^{-\gamma\tau\sqrt{3}} \right]
\end{multline}
where the internal temperature $T_\mathrm{int}$ is a measure of the planet's internal heat flux $\sigma T^4_\mathrm{int}$ -- the rate at which the planet cools -- $T_\mathrm{irr}$ a measure of the heat flux received from the star, $\tau$ the optical depth at IR wavelengths, and $\gamma = \kappa_\mathrm{vis}/\kappa_\mathrm{IR}$ the ratio between the opacity at visual (irradiated) and IR (outgoing) wavelengths. The parameter $f_\mathrm{irr}$ specifies the distribution of the incoming flux over the planet ($f_\mathrm{irr}=1/4$ for an equal distribution over the entire planet is used here). The irradiation temperature is defined $T_\mathrm{irr}=T_\star (R_\mathrm{p}/r_p)^2$ where $T_\star$ is the stellar (effective temperature), $R_p$ the planet radius and $r_p$ the distance to the host star. See \citet{Guillot2010} for further discussion.

Employing this relation between $T$ and $\tau$, the hydrostatic balance, and the ideal gas law we obtain the temperature and density as function of pressure. (In particular, for constant $\kappa_\mathrm{IR}$, as we will use here, we have $P=g_z \tau/\kappa_\mathrm{IR}$). \Fg{HJ-profile} provides the P-T and $\rho$-T structures resulting from the atmosphere model for our generic hot Jupiter model and GJ1214 b (parameters are discussed in \se{results} and listed in \Tb{parameters}). From the temperature a saturation vapor density ($P_\mathrm{sat}$) can be calculated. Vapor is expected to condense out when the partial pressure will exceeds $P_\mathrm{sat}$. Here, we re-express this condition in terms of a density: condensation occurs when $\rho_\mathrm{eq}<\rho_v$, where $\rho_\mathrm{eq} = m_v P_\mathrm{sat}/k_B T$, $\rho_v = x_v \rho$ the vapor density, $k_B$ Boltzmann constant, $x_v$ the mass concentration, and $m_v$ the molecular weight of the vapor species. 

Let us denote the vapor concentration below the cloud deck by $x_\mathrm{v,bot}$. Below the cloud deck we can expect that $x_\mathrm{v,bot}$ is constant. The vapor density below the cloud deck is then simply $\rho_v=x_\mathrm{v,bot}\rho_\mathrm{gas}$ (blue line in \fg{HJ-profile}). The height where $\rho_v=\rho_\mathrm{eq}$ (vertical dashed line) can be taken as the point where cloud formation starts -- the base of the cloud. Because of transport effects it is conceivable that the cloud will extend below this height, for example, heavy rain particles take time to evaporate. Conversely, the cloud deck could be located at a higher layer if cloud formation requires strong, super-saturated conditions. These considerations are automatically accounted for in our numerical model.

Above the cloud base the vapor concentration is expected to become less than $x_\mathrm{v,bot}$ (light blue line) due to cloud formation.
The surface (top) of the cloud is defined where the concentration of condensates $x_c=\rho_c/\rho_\mathrm{gas}$ is (close to) zero. The height where this occurs is not a priori known.

\subsection{Cloud transport model}
We model cloud transport of particles and vapor as advection-diffusion processes, solving equations like
\begin{equation}
    \frac{\partial\rho_i}{\partial t} +\nabla\cdot\mathcal{M}_i
    = \mathcal{S}_i
    \label{eq:rhoc-ti}
\end{equation}
where $\rho_i$ is the mass density of a certain species $i$, $t$ is time, and $\mathcal{M}_i$ the mass flux 
\begin{equation}
    \label{eq:M-flux}
    \mathcal{M}_i \equiv \rho_i \bm{v}_{\mathrm{sed},i} -\bm{\mathrm{K}} \rho_\mathrm{gas} \nabla x_i
\end{equation}
$x_i$ the mass concentration of species $i$, $v_{\mathrm{sed},i}$ the particle sedimentation velocity, $\bm{\mathrm{K}}$ the diffusion tensor, and $\rho_\mathrm{gas}$ the gas density. In this work, we consider only vertical ($z$) transport, implying that only one velocity component and one diffusion element ($K_\mathrm{zz}$) remain ($\bm{v}_\mathrm{sed}=0$ for a vapor species). The RHS of \eq{rhoc-ti}, $\mathcal{S}_i$, specifies source terms arising from deposition (condensation), sublimation (evaporation) or nucleation, depending on the species $i$. The sedimentation velocity is obtained by equating the aerodynamic drag force with the planet's gravity, $v_\mathrm{sed,p} = g_z t_\mathrm{stop}$ where $t_\mathrm{stop}$ encapsulates the aerodynamic properties of the cloud particle. In general, the gas drag force is non-linear in particle-gas velocity $v_\mathrm{sed}$ (see e.g., \citealt{Whipple1972}) and $t_\mathrm{stop}$ must be found by iteration. However, for small particles $t_\mathrm{stop}$ becomes independent of velocity. In particular, for the parameters of our model cloud, the gas drag law obeys the \citet{Epstein1924} regime (free molecular flow) for which
\begin{equation}
    t_\mathrm{stop-Epstein} = \frac{a_p\rho_\bullet}{v_\mathrm{th} \rho_\mathrm{gas}}
    \label{eq:tEpstein}
\end{equation}
where $a_p$ is the radius of the grain, $\rho_\bullet$ its internal density, $v_\mathrm{th}=\sqrt{8k_B T/\pi m_\mathrm{gas}}$ the thermal velocity of the gas. %, and $C_D$ the drag coefficient. 
The Epstein drag law applies in the free molecular flow regime, $a_p<\frac{9}{4}l_\mathrm{mfp}$, where $l_\mathrm{mfp} = m_\mathrm{gas}/(\sqrt{2} \rho_\mathrm{gas} \sigma_\mathrm{mol})$
is the mean free path. 

We employ the following assumptions:
\begin{enumerate}
    \item The medium consist of three components -- nuclei (n), condensates (c), and vapor (v). Only a single cloud species is considered. Any gas-gas or gas-grain chemistry is not accounted for. 
    \item The model is plane parallel; only the vertical dimension ($z$) is modeled and the only relevant diffusion coefficient is $K_\mathrm{zz}$.%The equations are solved under the plane parallel assumption ($\nabla \rightarrow \partial/\partial z$).
    \item The cloud model is in steady state, $\partial/\partial t =0$. This implies that
        \begin{equation}
            \mathcal{M}_v = -\mathcal{M}_c
            \label{eq:mass-equi}
        \end{equation}
        at any location.
    \item Nucleation is parametrized in the form of a log-normal profile with height (pressure)
\begin{equation}
    \mathcal{S}_n 
    = \rho_\mathrm{gas} g_z \frac{\dot{\Sigma}_N}{\sigma_\ast P \sqrt{2\pi}}
    \exp \left[ -\frac{1}{2\sigma_\ast^2} \left( \log \frac{P}{P_\ast} \right)^2 \right]
    \label{eq:S-nucl}
\end{equation}
where $\dot{\Sigma}_N$, $P_\ast$ and $\sigma_\ast$ respectively indicate the integrated nuclei production rate, the characteristic height where the nuclei are deposited, and the width of the distribution. %The chemical makeup of the nuclei is assumed irrelevant as they barely contribute to the total mass of the particles.
    \item At any layer, the characteristic particle mass $m_p$ is obtained taking the ratio of the total solid density of the particles (the density of condensates $\rho_c = x_c \rho_\mathrm{gas}$ plus the density of nuclei $x_n \rho_\mathrm{gas}$) to the particle number density $n_p$. In our model, the particle number density follows from the nuclei number density. In the case without coagulation any particle will contain only one nuclei, $n_p=n_n$. (Below, in \se{coagulation} this assumption will be relaxed, accounting for coagulation effects.) Hence
\begin{equation}
    m_p = \frac{(x_c+x_n)\rho_\mathrm{gas}}{n_p}
    \approx \frac{x_c \rho_\mathrm{gas}}{n_n}
    = \frac{x_c m_{n0}}{x_n}
    \label{eq:pprop}
\end{equation}
where $m_{n0}$ is the mass of a single nuclei. In \eq{pprop} the second step assumes that the condensates dominate the mass and the last step employs the single nuclei per cloud particle assumption: $n_n m_{n0}=x_n\rho_\mathrm{gas}$.  From the characteristic particle mass $m_p$ a characteristic grain radius $a_p$ follows, assuming that the grains are spherical. A grain size distribution is not accounted for, but it may be reconstructed from $a_p$.  In addition, $a_p$ changes with height through nucleation, condensation, evaporation, and coagulation. The grain radius $a_p$ in turn determines the sedimentation velocity $v_\mathrm{sed,p}$ of the particles.%by balancing the gas drag law with gravitational force, $F_\mathrm{aero}(v_\mathrm{sed,p}) = g_z$. 
    \item We take the diffusivity ($K_\mathrm{zz}$) equal for vapor and particles and independent of height. (These assumptions are easily relaxed, though).
\end{enumerate}

\noindent
We then obtain the following set of ordinary equations specifying the evolution of the condensate, nuclei, and vapor:
\begin{subequations}
    \begin{equation}
    \label{eq:new1}
    \frac{\partial \mathcal{M}_c}{\partial z} = \mathcal{S}_c \\
    \end{equation}
    \begin{equation}
        \label{eq:Mn}
    \frac{\partial \mathcal{M}_n}{\partial z} = \mathcal{S}_n \\
    \end{equation}
    \begin{equation}
        \frac{\partial x_c}{\partial z} = \frac{x_c v_\mathrm{sed,p}}{K_\mathrm{zz}} -\frac{\mathcal{M}_c}{K_\mathrm{zz}\rho_\mathrm{gas}} \\
    \end{equation}
    \begin{equation}
        \label{eq:xn}
        \frac{\partial x_n}{\partial z} = \frac{x_n v_\mathrm{sed,p}}{K_\mathrm{zz}} -\frac{\mathcal{M}_n}{K_\mathrm{zz}\rho_\mathrm{gas}} \\
    \end{equation}
    \begin{equation}
    \label{eq:new5}
    \frac{\partial x_v}{\partial z} = -\frac{\mathcal{M}_v}{K_\mathrm{zz}\rho_\mathrm{gas}} %= +\mathcal{M}_c/K_v\rho_\mathrm{gas}
    \end{equation}
\end{subequations}
where $\mathcal{M}_v$ is given by \eq{mass-equi}, $\mathcal{S}_n$ by \eq{S-nucl}, and the particle properties follow from $x_c$ and $x_n$ as described above. 

The condensation rate $\mathcal{S}_c$ is given by
\begin{equation}
\mathcal{S}_c = f_\mathrm{stick} (x_v \rho_\mathrm{gas} -\rho_\mathrm{eq}) \times \min\left[ \pi a_p^2 v_{\mathrm{th},v}n_p; 4\pi a_p D_i n_p \right].
    \label{eq:source}
\end{equation}
where $D_i$ the diffusivity and $f_\mathrm{stick}$ a sticking probability, here taken unity, and $\rho_\mathrm{eq}=m_v P_\mathrm{sat}/k_B T$ the equilibrium (or saturation) density.  \Eq{source} combines the free molecular flow (vapor molecules travel on ballistic trajectories on the scale of the particle) and the diffusion-limited regimes \citep{WoitkeHelling2003,YauRogers1996}. In \eq{source} we have not accounted for the (liberated) latent heat of condensation.

For the diffusivity we follow \citet{WoitkeHelling2003}, after \citet{Jeans1967}, and write
\begin{equation}
    D_i = \frac{k_B T}{3 P_\mathrm{gas} \sigma_\mathrm{com}} v_\mathrm{th,red}.
\end{equation}
This equation describes diffusion of a quantity in a two component medium of vapor and hydrogen gas.  The reduced thermal velocity $v_\mathrm{th,red}$  is taken equal to the mean gas thermal velocity ($v_\mathrm{th,red}=v_\mathrm{th}$) and $\sigma_\mathrm{com}$ is the combined cross section. We take, somewhat arbitrarily, $\sigma_\mathrm{com}=8\times10^{-15}\, \mathrm{cm^2}$. 

\begin{table*}
    \centering
    \caption{Cloud and atmospheric parameters for the generic hot-Jupiter and GJ1214 b.}
    \begin{tabular}{lrrlp{8cm}}
        \hline
        \hline
        Symbol                  & \multicolumn{2}{c}{(\textbf{default}) Value} & Unit                                      & Description \\
        \cline{2-3} \\[-1em]
        &   Generic HJ      & GJ1214 b\tablefootmark{a}          \\
        \hline \\[-1em]
        species                 & \mgsio            & KCl               &                                       & cloud species \\
        $\dot{\Sigma}_n$        & $10^{-19}$, $\bm{10^{-15}}$, $10^{-11}$ &   & $\mathrm{g\,cm^{-2}\,s^{-1}}$   & nucleation rate \\
        $\gamma$                & $0.158$           & $0.038$           &                                       & opacity ratio visual and IR wavelengths (\eq{Guillot}) \\ 
        $\kappa_\mathrm{IR}$    & $0.3$             & $0.03$            & $\mathrm{cm^2\,g^{-1}}$               & IR opacity \\ 
        $\rho_\bullet$          & $2.8$             & $2.8$             & $\mathrm{g\,cm^{-3}}$                 & particle internal density \\ 
        $\sigma_\ast$           & $0.2$             &                   &                                              & width of nucleation profile (\eq{S-nucl}).  \\
        $\sigma_\mathrm{com}$   & $8\times10^{-15}$ &                   & $\mathrm{cm^2}$                       & combined (vapor and gas) molecular cross section \\
        $\sigma_\mathrm{mol}$   & $2\times10^{-15}$ &                   & $\mathrm{cm^2}$                       & molecular cross section (gas)  \\
        $K_g, K_p$              & $10^6,\bm{10^8},10^{10}$ & $10^8$     & $\mathrm{cm^2\,s^{-1}}$               & particle and gas diffusivity \\
        $M_\mathrm{planet}$     & $1$               & $0.0206$          & $M_\mathrm{J}$                        & planet mass  \\
        $P_\ast$                & $60$              & $10^4$            & $\mathrm{dyn\,cm^{-2}}$               & reference height for the nucleation profile (\eq{S-nucl}) \\
        $R_\mathrm{pl}$         & $1.087$           & $0.244$           & $R_J$                                             & planet radius \\
        $R_\star$               & $1$               & $0.2064$          & $R_\odot$                                         & stellar radius   \\
        $T_\star$               & $5778$            & $3026$            & K                                                 & stellar effective temperature   \\
        $T_\mathrm{int}$        & $500$             & $60$              & K                                                 & internal temperature \\
        $a_n$                   & $0.001$           &                   & $\mu\mathrm{m}$                                   & particle nucleation radius \\ 
        $f_\mathrm{irr}$        & $0.25$            &                   &                                       & heat distribution factor \\ 
        $f_\mathrm{stick}$      & $1.0$             &                   &                                       & vapor sticking probability \\ 
        $g_z$                   & $2192$            & $893$             & $\mathrm{cm\,s^{-2}}$                             & gravitational acceleration \\
        $m_\mathrm{gas}$        & $2.34$            &                   & $m_\mathrm{H}$                                    & mean molecular weight (gas) \\ 
        $m_v$                   & $34.67$           & $74.45$           & $m_\mathrm{H}$                                    & mass vapor species \\
        $r_p$                   & $0.05$            & $0.0143$          & $\mathrm{au}$                                     & distance to star \\
        $x_\mathrm{v,bot}$      & $3\times10^{-3}$  & $3\times10^{-4}$  &                                                   &vapor mass concentration at/below cloud base \\
        \hline
        \hline
    \end{tabular}
    \tablefoot{
    \tablefoottext{a}{Empty entries indicate the value listed in the generic hot Jupiter column is used.}
}
    \label{tab:parameters}
\end{table*}
\begin{table*}[tb]
    \caption{Table of output quantities.}
    \centering
 \begin{tabular}{lllllllllll}
\hline
\hline
planet             & coagulation             & $K_\mathrm{zz}$  & $\dot\Sigma_n$             & $\mathcal{M}_\mathrm{c,max}$             & $P_{\tau=1}$            & $\tau_\mathrm{z,tot}$            & $a_\mathrm{max}$  \\
            &             & $[\mathrm{cm^2\,s^{-1}}]$             & $[\mathrm{g\,cm^{-2}\,s^{-1}}]$             & $[\mathrm{g\,cm^{-2}\,s^{-1}}]$             & [bar]            &             & $[\mu\mathrm{m}]$  \\
\hline
generic HJ  & x           & {$10^{10}$}  & {$10^{-11}$} & {$-1.8\times10^{-6}$}  & {$4.3\times10^{-7}$} & {$3.0\times10^{3}$} & {$0.051$}    \\
            & \checkmark  & {$10^{10}$}  & {$10^{-11}$} & {$-1.7\times10^{-6}$}  & {$9.5\times10^{-7}$} & {$740$}      & {$0.24$}     \\
            & x           & {$10^{10}$}  & {$10^{-15}$} & {$-1.6\times10^{-6}$}  & {$8.9\times10^{-6}$} & {$140$}      & {$1.0$}      \\
            & \checkmark  & {$10^{10}$}  & {$10^{-15}$} & {$-1.8\times10^{-6}$}  & {$1.0\times10^{-5}$} & {$73$}       & {$2.4$}      \\
            & x           & {$10^{10}$}  & {$10^{-19}$} & {$-1.3\times10^{-6}$}  & {$2.6\times10^{-4}$} & {$5.0$}      & {$20$}       \\
            & \checkmark  & {$10^{10}$}  & {$10^{-19}$} & {$-1.3\times10^{-6}$}  & {$2.7\times10^{-4}$} & {$3.9$}      & {$26$}       \\
            & x           & {$10^{8}$}   & {$10^{-11}$} & {$-1.9\times10^{-8}$}  & {$3.5\times10^{-6}$} & {$1.3\times10^{4}$} & {$0.012$}    \\
            & \checkmark  & {$10^{8}$}   & {$10^{-11}$} & {$-1.8\times10^{-8}$}  & {$1.9\times10^{-5}$} & {$29$}       & {$3.9$}      \\
            & x           & {$10^{8}$}   & {$10^{-15}$} & {$-1.9\times10^{-8}$}  & {$3.6\times10^{-5}$} & {$460$}      & {$0.25$}     \\
            & \checkmark  & {$10^{8}$}   & {$10^{-15}$} & {$-1.8\times10^{-8}$}  & {$4.6\times10^{-5}$} & {$22$}       & {$4.1$}      \\
            & x           & {$10^{8}$}   & {$10^{-19}$} & {$-1.8\times10^{-8}$}  & {$2.1\times10^{-3}$} & {$4.6$}      & {$5.2$}      \\
            & \checkmark  & {$10^{8}$}   & {$10^{-19}$} & {$-1.8\times10^{-8}$}  & {$2.1\times10^{-3}$} & {$2.5$}      & {$8.7$}      \\
            & x           & {$10^{6}$}   & {$10^{-11}$} & {$-1.9\times10^{-10}$} & {$1.0\times10^{-4}$} & {$6.2\times10^{4}$} & {$2.7\times10^{-3}$} \\
            & \checkmark  & {$10^{6}$}   & {$10^{-11}$} & {$-1.9\times10^{-10}$} & {$6.6\times10^{-4}$} & {$1.7$}      & {$1.4$}      \\
            & x           & {$10^{6}$}   & {$10^{-15}$} & {$-1.9\times10^{-10}$} & {$2.2\times10^{-4}$} & {$450$}      & {$0.058$}    \\
            & \checkmark  & {$10^{6}$}   & {$10^{-15}$} & {$-1.9\times10^{-10}$} & {$3.8\times10^{-3}$} & {$1.6$}      & {$1.3$}      \\
            & x           & {$10^{6}$}   & {$10^{-19}$} & {$-1.9\times10^{-10}$} & {$5.6\times10^{-3}$} & {$1.3$}      & {$1.2$}      \\
            & \checkmark  & {$10^{6}$}   & {$10^{-19}$} & {$-1.9\times10^{-10}$} & {$5.8\times10^{-3}$} & {$0.86$}     & {$1.7$}      \\
GJ1214 b    & x           & {$10^{8}$}   & {$10^{-11}$} & {$-3.2\times10^{-8}$}  & {$5.6\times10^{-6}$} & {$2.7\times10^{4}$} & {$0.015$}    \\
            & \checkmark  & {$10^{8}$}   & {$10^{-11}$} & {$-3.0\times10^{-8}$}  & {$2.0\times10^{-4}$} & {$810$}      & {$1.3$}      \\
            & x           & {$10^{8}$}   & {$10^{-15}$} & {$-3.1\times10^{-8}$}  & {$1.2\times10^{-4}$} & {$1.2\times10^{3}$} & {$0.30$}     \\
            & \checkmark  & {$10^{8}$}   & {$10^{-15}$} & {$-3.0\times10^{-8}$}  & {$4.1\times10^{-4}$} & {$250$}      & {$2.5$}      \\
            & \checkmark  & {$10^{8}$}   & {$10^{-19}$} & {$-2.7\times10^{-8}$}  & {$3.0\times10^{-3}$} & {$16$}       & {$17$}       \\
            & x           & {$10^{8}$}   & {$10^{-19}$} & {$-2.9\times10^{-8}$}  & {$2.5\times10^{-3}$} & {$44$}       & {$6.0$}      \\
\hline
\end{tabular}
    \tablefoot{The first four columns list the input parameters: the planet (see \tb{parameters} for parameters), whether coagulation is included or not, the diffusivity $K_\mathrm{zz}$, and the nuclei production rate $\dot{\Sigma}_n$. The model calculates a steady state cloud and we list: the peak mass flux (intensity of the rain) $\mathcal{M}_\mathrm{c,max}$, the pressure level where the geometrical transmission optical depth reaches unity $P_\mathrm{\tau=1}$, the total geometrical vertical optical depth of the cloud $\tau_\mathrm{z,tot}$, and the maximum grain radius $a_\mathrm{max}$.}
    \label{tab:output}
\end{table*}
\subsection{Adding coagulation}
\label{sec:coagulation}
Coagulation among the cloud particles will decrease their number density $n_p$, relaxing the identity $n_p=n_n$, while leaving unaffected the mass concentration of nuclei. 
That is, coagulation will result in $n_p < x_n\rho_\mathrm{gas} /m_\mathrm{n0}$. Within the above framework, it is possible to include coagulation among the cloud particles by adding two additional equations, describing $n_p$:
\begin{subequations}
    \label{eq:coag1}
    \begin{equation}
        \label{eq:coag1a}
        \frac{\partial \mathcal{N}_p}{\partial z} = \frac{\mathcal{S}_n}{m_\mathrm{n0}} - \frac{n_p}{t_\mathrm{coag}}
    \end{equation}
    \begin{equation}
        \label{eq:coag1b}
        \frac{\partial c_p}{\partial z} = c_p v_\mathrm{sed,p}/K_p -\mathcal{N}_p/K_p n_\mathrm{gas}
    \end{equation}
\end{subequations}
where $\mathcal{N}_p$ is the particle number flux and $t_\mathrm{coag}$ is the coagulation timescale, $c_p = n_p/n_\mathrm{gas}$ the particle concentration (by number), and $n_\mathrm{gas}$ the gas number density. The coagulation time includes contributions from differential settling ($\Delta v$) and Brownian motion. In terms of the coagulation rate ($dn_p/dt = - n_p/t_\mathrm{coag}$) these can be added:
\begin{equation}
    \label{eq:tcoag}
    t_\mathrm{coag}^{-1}
    = \frac{1}{2} n_p \pi (2a_p)^2 \Delta v
    +\frac{1}{2} 4\pi \min (v_\mathrm{BM}a_p, D_p) a_p n_p
\end{equation}
where $v_\mathrm{BM}=\sqrt{16 k_B T/\pi m_p}$ for equal mass particles, $D_p = k_B T/6\pi \eta a_p$ (Stokes–Einstein equation), $\eta = \nu_\mathrm{mol} \rho_\mathrm{gas}$ the dynamic viscosity, $\nu_\mathrm{mol}=0.5 l_\mathrm{mfp} v_\mathrm{th}$ the molecular viscosity \citep{ChapmanCowling1970}, and $\Delta v$ is the relative velocity between the cloud particles. The factor $\frac{1}{2}$ prevents double counting. Following \citet{KrijtEtal2016} and \citet{SatoEtal2016} it is appropriate to take $\Delta v = 0.5v_\mathrm{sed}$ when the coagulation is driven by sedimentation. For identical particles having the same aerodynamical properties $\Delta v=0$, but in reality a distribution in aerodynamical properties always ensures that $\Delta v \sim v_\mathrm{sed}$ \citep{OkuzumiEtal2011}.

Adding these equations would bring the total number of equations to solve to seven. However, when we assume (correctly) that the nuclei mass is insignificant, $x_n \ll x_c$, there is no need to follow the nuclei mass density $x_n$. \Eq{coag1a} and \eq{coag1b} then replace \eq{Mn} and \eq{xn}. To keep the expressions in units of mass concentrations (like $x$) and mass flux (like $\mathcal{M}$), we transform \eq{coag1} by defining:
\begin{subequations}
    \begin{equation}
        \tilde{\mathcal{M}}_n = m_{n0} \mathcal{N}_p
    \end{equation}
    \begin{equation}
        \tilde{x}_n = n_p m_{n0} /\rho_\mathrm{gas}
    \end{equation}
\end{subequations}
In terms of these new variables, \eq{coag1} read
\begin{subequations}
    \label{eq:coag2}
    \begin{equation}
        \frac{\partial \tilde{\mathcal{M}}_n}{\partial z} = \mathcal{S}_n - \frac{\tilde{x}_n \rho_\mathrm{gas}}{t_\mathrm{coag}} \\
    \end{equation}
    \begin{equation}
        \frac{\partial \tilde{x}_n}{\partial z} = \tilde{x}_n v_\mathrm{sed,p}/K_p -\mathcal{\tilde{M}}_n/K_p \rho_\mathrm{gas}
    \end{equation}
\end{subequations}
These are identical to \eq{Mn} and \eq{xn}, except for the term involving $t_\mathrm{coag}$. In runs including coagulation we simply use these equations to follow the number density of nuclei ($n_n$ or $\tilde{x}_n$). The nuclei mass density ($x_n$) is not followed, but this is justified since it is in any case negligible compared to the mass density of the condensate ($x_c$) and therefore bears no influence on the physical properties of the cloud particles.

\begin{figure*}[bt]
    \includegraphics[width=\textwidth]{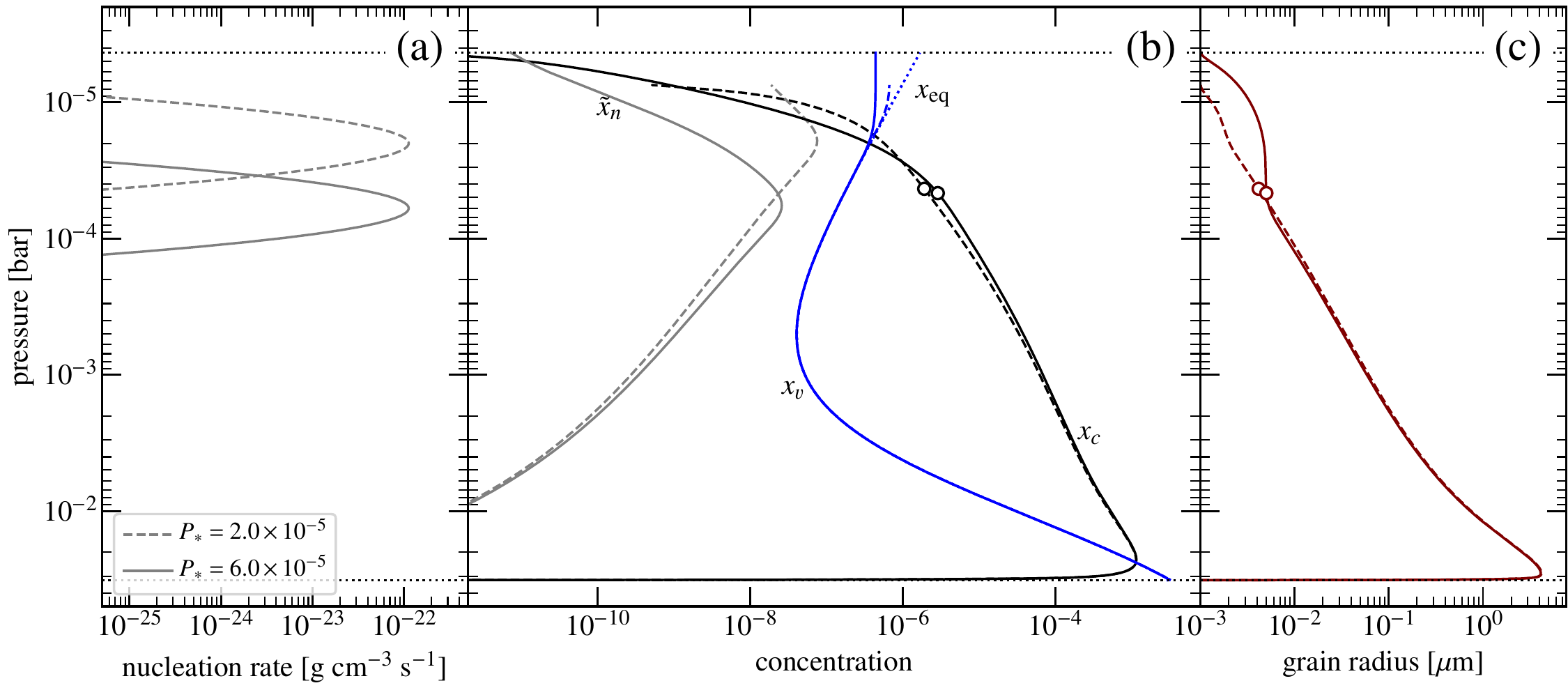}
    \caption{Results of the standard model ($K=10^8\,\mathrm{cm^2\,s^{-1}}$; $\dot\Sigma_n=10^{-15}\,\mathrm{g\,cm^{-2}\,s^{-1}}$). (a) Nucleation rate $\mathcal{S}_n$, which follows a log-normal distribution around a reference pressure $P_\ast$. The dotted horizontal lines indicate the bottom and top of our computational domain. (b) The concentrations of nuclei ($\tilde{x}_n$), condensates ($x_c$), and vapor ($x_v$). The equilibrium concentration corresponding to the saturation pressure is also plotted ($x_\mathrm{eq}$) but it virtually coincides with $x_v$. (c) The characteristic grain radius $a_p$. The dashed lines correspond to a model where the nuclei are inserted at a higher layer than the standard. The open circles correspond to the depth where the geometrical transmission optical depth $\tau_\mathrm{trans}$ equals 1.}
    \label{fig:standard}
\end{figure*}
\begin{figure*}
    \centering
    \includegraphics[width=\textwidth]{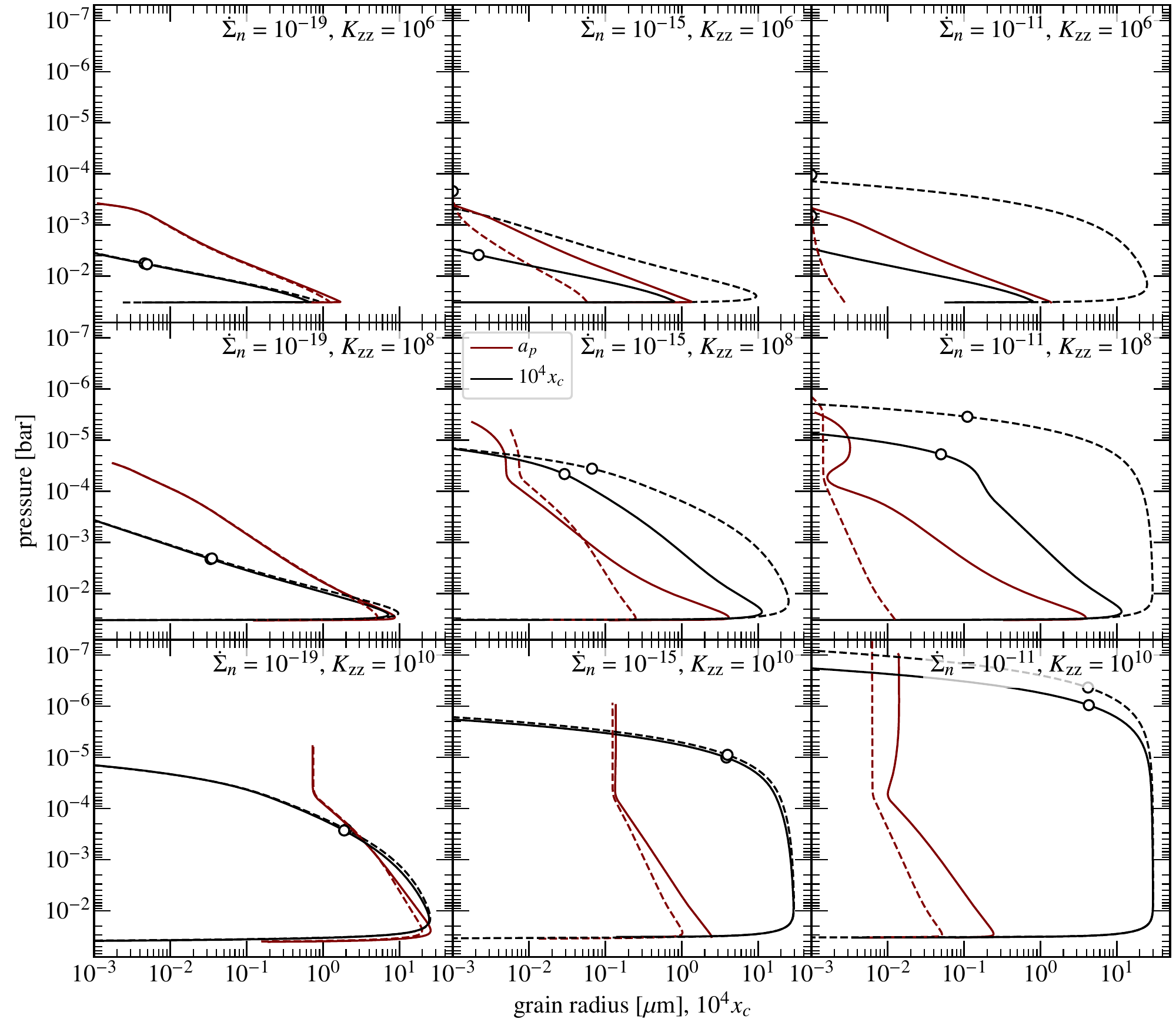}
    \caption{Cloud profiles. The concentration of cloud particles (black) and the characteristic particle size (dark red; shared x-axis) against pressure, plotted for combinations of diffusivities $K_\mathrm{zz}$ and nucleation rates $\dot{\Sigma}_n$ (panels). 
    The grain radius of particles in models without coagulation is shown by the dashed dark curve. The height where the transition optical depth reached unity ($\tau_\mathrm{trans}=1$) is indicated by the circle.}
    \label{fig:collage}
\end{figure*}

\subsection{Boundary conditions and solution technique}
\Eqs{new1}{new5} constitute a system of five first order, ordinary differential equations and five unknowns ($x_c, x_n, x_v, \mathcal{M}_c, \mathcal{M}_n$). Therefore, five boundary conditions are necessary. We specify boundary conditions at the bottom and the top of the domain. At the top of the cloud ($z=z_\mathrm{top}$) we demand that the condensate flux vanishes:
\begin{equation}
    \mathcal{M}_c(z_\mathrm{top}) = 0
    \label{eq:BCtop}
\end{equation}
and that the nuclei flux equals
\begin{equation}
    \mathcal{M}_n(z_\mathrm{top}) = -\int_{z_\mathrm{top}}^\infty \mathcal{S}_n dz
    \label{eq:BCtop1}
\end{equation}
while at the base of the cloud we put constraints on the mass concentrations:
\begin{equation}
    x_n(z_\mathrm{bot}) = x_c(z_\mathrm{bot})=0;\quad x_v(z_\mathrm{bot})=x_\mathrm{v,bot}. %;\quad x_a(z_\mathrm{bot}) = x_a^\ast
    \label{eq:BCbot}
\end{equation}
The condition $x_c=0$ reflects that at the base of the cloud the temperature has become high enough for all the condensates to be evaporated. The vapor concentration at the cloud base ($x_\mathrm{v,bot}$) is an input. The nuclei boundary condition $x_n=0$ strictly only holds when the nuclei are also made of \mgsio, such that they would also evaporate. But this is not necessarily the case. Formally, we should extend the systems of equations describing evaporation of the nuclei species, which is a rather cumbersome extension of the model. Alternatively, we could introduce a free parameter for $n_n(z_\mathrm{bot})$. But we found its effects rather insignificant as long as it is not too large. Hence, we considered the simple choice of a zero concentration nuclei boundary condition is preferable above an (arbitrary) specification of the nuclei seed.

Since conditions are placed on both the upper and the lower boundary, \eqs{new1}{new5} represent a boundary value problem (BVP). This BVP is solved using the \texttt{solve\_bvp} function from python's \texttt{SciPy} module \citep{AscherEtal1994,KierzenkaShampine2001,ShampineEtal2006}. These codes requires an initial ``guess'' for the solution, which must be sufficiently close to the actual solution. Otherwise, convergence is not guaranteed. This represents a problem since the actual solution is of course unknown.

We therefore resort to an iterative approach, introducing a parameter $\epsilon$ that is added to the condensation rate $\mathcal{S}_c$ and the nucleation rate $\mathcal{S}_n$. Hence, $\epsilon=0$ corresponds to the cloud-free solution ($x_v=x_\mathrm{v,bot}$, $x_c=0$, $\mathcal{M}_c=0$). Then, a very small $\epsilon$, for example $\epsilon=10^{-8}$, will give a solution that will be close to the known (cloud-free) solution that \texttt{solve\_bvp} is able to solve. This new solution (with $\epsilon=10^{-8}$) then provides the guess for the next iteration, where $\epsilon$ is larger. We progressively increase $\epsilon$ until $\epsilon=1$, with which the desired cloud profiles are obtained.

A similar iterative approach can be designed for the boundaries of the domain. Although the bottom boundary is given by the $\rho_\mathrm{v,bot}$ constraint,\footnote{The lower boundary may deviated from the $\rho_\mathrm{v,bot}=\rho_\mathrm{eq}$ condition when transport timescales are shorter than evaporation times, e.g., when the particles have become large and settle quickly.} the upper boundary is in principle open, as diffusion always allows some particles to be transported to the very upper regions. As a final step, we therefore adjust the boundaries of the domain, searching for a solution where $x_c$ stays positive in the entire domain, while $x_c$ near the boundary is a very tiny fraction (e.g., $10^{-8}$) of its peak value.

With these incremental approach of ``relaxing'' to the solution, \texttt{solve\_bvp} is still computational efficient. The 24 runs listed in \Tb{output} took an average of 17 seconds to complete on a modern desktop PC, with the slowest one requiring 25 seconds.

\section{Physical structure}
\label{sec:results}
In this section we present the outcome of the cloud model in terms of its physical structure: the concentration and properties of the cloud particles.  In \se{physical} we consider a generic hot Jupiter planet, whereas in \se{GJ1214b} we apply our model toward GJ1214 b to compare our results to previous findings.

\subsection{Hot-Jupiter \mgsio clouds}
\label{sec:physical}

We consider a generic hot Jupiter planet situated at a distance of 0.05\,au around a solar-like star. We consider \mgsio as our cloud species, for which we use the saturation pressure of \citet{AckermanMarley2001}
\begin{equation}
    \label{eq:Psat}
    P_\mathrm{sat}
    = 1.04\times10^{17} \exp \left[ -\frac{58\,663}{T} \right] \mathrm{dyn\,cm}^{-2}.
\end{equation}
Because \mgsio does not exist in vapor phase, it would be erroneous to consider taking the molecular weight of \mgsio (100.4 $m_H$) for $m_v$. Instead, we consider an effective vapor mass, which is given by the constituents from which \mgsio forms. Typically, \mgsio falls apart into three molecules \citep{HellingEtal2008}. We therefore simply take $m_v=m_\textrm{\mgsio}/3=33.5 m_H$.
For the atmospheric parameters, we adopt parameters similar values as \citet{LineEtal2013}, see \tb{parameters}. An internal temperature of $T_\mathrm{int}=500$ K is used and an atmosphere IR-opacity of $0.3\,\mathrm{cm^2\,g^{-1}}$. The higher IR-opacity crudely reflects the appearance of clouds; the model does presently not treat (thermal) feedback of the clouds on the profiles consistently.  We have verified that changing these parameters does not affect the conclusions of this work. The corresponding atmospheric physical structure was shown in \fg{HJ-profile}.

The outcome of the cloud model for the parameters listed in \Tb{parameters} is presented in \fg{standard} for the default model. In \fg{collage} we take eight other parameter combinations of $K_\mathrm{zz}$ and $\dot\Sigma_n$, crudely corresponding what has been used in literature studies \citep[e.g.,][]{KawashimaIkoma2018}. Several output quantities of the runs are further listed in \Tb{output}. The intensity of the rain is characterized by the mass flux parameter $\mathcal{M}_\mathrm{c}$ whose peak value is listed. A higher $\mathcal{M}_\mathrm{c,max}$ reflects more vigorous mass transport; this parameter hence correlates with the diffusivity $K_\mathrm{zz}$. For reference, a value of $\mathcal{M}_\mathrm{c}=10^{-7}\,\mathrm{g\,cm^{-2}\,s^{-1}}$ amounts to a \mgsio precipitation of 11 $\mathrm{mm\,yr^{-1}}$.  We calculate both the vertical optical depth %(0.44 inch\,yr$^{-1}$)
\begin{equation}
    \label{eq:tau-z}
    \tau_z(z) = \int_z^{z_\mathrm{top}} n_p(z') \pi a_p^2(z') dz'
\end{equation}
as well as the transmission optical depth in the geometrical limit
\begin{equation}
    \label{eq:trans-z}
    \tau_\mathrm{trans}(z) = \int_z^{z_\mathrm{top}} n_p(z') \pi a_p^2(z') \sqrt{\frac{2R}{(z'-z)}} dz'
\end{equation}
that is, the optical depth corresponding from the line perpendicular to height $z$. 
In \Tb{output} the total geometrical optical depth refers to $\tau_z$ as measured from the base of the cloud whereas the pressure level where $\tau$ reaches unity ($P_\mathrm{\tau=1}$) refers to the transmission optical depth $\tau_\mathrm{trans}$. The latter quantity is more meaningful in the context of transmission spectra. These geometrical values only serve as a crude guide as opacities are not often close to their geometrical limit (especially for small particles). Proper simulated spectra are calculated in \se{spectra}. Finally, we list the peak radius of the condensate particles, $a_\mathrm{max}$.

\Fg{standard} presents profiles of nucleation rate, concentrations of vapor condensates and nuclei and grain size for the standard model ($K_\mathrm{zz}=10^8\,\mathrm{cm^2\,s^{-1}}$, $\dot\Sigma_n=10^{-15}\,\mathrm{g\,cm^{-2}\,s^{-1}}$; the central panel of \fg{collage} corresponds to \fg{standard}). Coagulation is included.  
Note the steep but continuous transition from cloudy to cloud-free near the bottom of the cloud. This is caused by the steep increase in the equilibrium density (\fg{HJ-profile}). Several factors regulate the extent of the cloud. The first is the location where the nuclei form, which is given in \fg{standard}a. Recall that the nuclei production profile $\mathcal{S}_n$ (\eq{S-nucl}) is characterized by three parameters: $P_\ast$, $\sigma_\ast$ and $\dot\Sigma_n$. In \fg{standard} we also present a case where the nuclei are released at a higher height ($P_\ast$ is decreased by a factor three; dashed curves). Increasing the height where the nuclei are released does not much affect the profiles deeper in the atmosphere. In both cases cloud particles readily consume the vapor locally, whereas transport and coagulation act on larger (time)scales. However, there may be some observational consequences as the grain size around $\tau_\mathrm{trans}=1$ is affected.

Comparing \fg{standard}a and b, it can be seen that the height where nuclei are injected is also the height where the concentration of nuclei ($\tilde{x}_n$) peaks. Below this height $\tilde{x}_n$ decreases because particles' velocity speeds up due to their growth by condensation. 
The ratio of $x_c$ and $\tilde{x}_n$ determines the size of the particles, which increases for our standard model to $4\,\mu\mathrm{m}$ just above the cloud base (\fg{standard}c). In the upper regions, the particle radius levels off at $\approx$0.005 $\mu$m, several factors larger than the nucleation radius. Grains tend to be somewhat smaller and more abundant in the model where the nuclei are injected at a larger height, because they accrete less vapor before settling down (\fg{standard}b and c). Although at these heights the density of \mgsio is rather low, the larger grain size may be of some observational importance for the transmission spectra, especially concerning the Rayleigh scattering at optical wavelengths.

More important in regulating the cloud thickness is the eddy diffusivity $K_\mathrm{zz}$. A larger $K_\mathrm{zz}$ implies that more vapor is transported upwards and that more (small) particles can be found above the nuclei injection height. This is illustrated in \fg{collage} where we vary the diffusivity (rows) and the total nucleation rate (columns). Clearly, larger diffusivity results in denser and thicker clouds; particles are uplifted to higher regions and more vapor is being transported from below the cloud deck. In the limit where the transport becomes dominated by diffusion, we can expect the concentration of condensates $x_c$ to be identical to the concentration of the vapor at the base of the cloud, $x_\mathrm{v,bot}=3\times10^{-3}$. This explains the boxy cloud profile of the $\dot\Sigma_n=10^{-11}$, $K_\mathrm{zz}=10^{10}$ model (bottom right panel of \fg{collage}).

Finally, we find that the cloud thickness increases with the nucleation rate $\dot\Sigma_n$ 
(\fg{collage}). A higher $\dot\Sigma_n$ tends to reduce the grain size, since the total amount of vapor on a given grain is smaller when there are more of them. Since the grain size is a directly observational property, the nucleation model is therefore an essential part to any cloud model. 

We also conducted runs without coagulation, in order to isolate its effects. These are presented by the dashed lines in \fg{collage}. Clearly, coagulation does not affect the low $\dot\Sigma_n$ runs. The growth of particles in these runs is entirely due to condensation. Coagulation is more important at higher nucleation rates and for lower $K_\mathrm{zz}$; the former because there is a larger surface available and the latter because the grains are more concentrated. The differences between the no-coagulation and coagulation runs are the greatest in the $(\dot\Sigma_n,K_\mathrm{zz})=(10^{-11},10^6)$ run (top right). The no-coagulation run is characterized by extremely small particles (similar to the nucleation size) and the total geometrical optical depth of the cloud reaches values above $10^4$ (see \Tb{output}). Including coagulation, however, greatly increases the grain size, reducing the cloud vertical geometrical optical depth by over a factor of $10^4$! 

\subsection{GJ1214 b KCl clouds}
\label{sec:GJ1214b}
\begin{figure}
    \centering
    \includegraphics[width=\columnwidth]{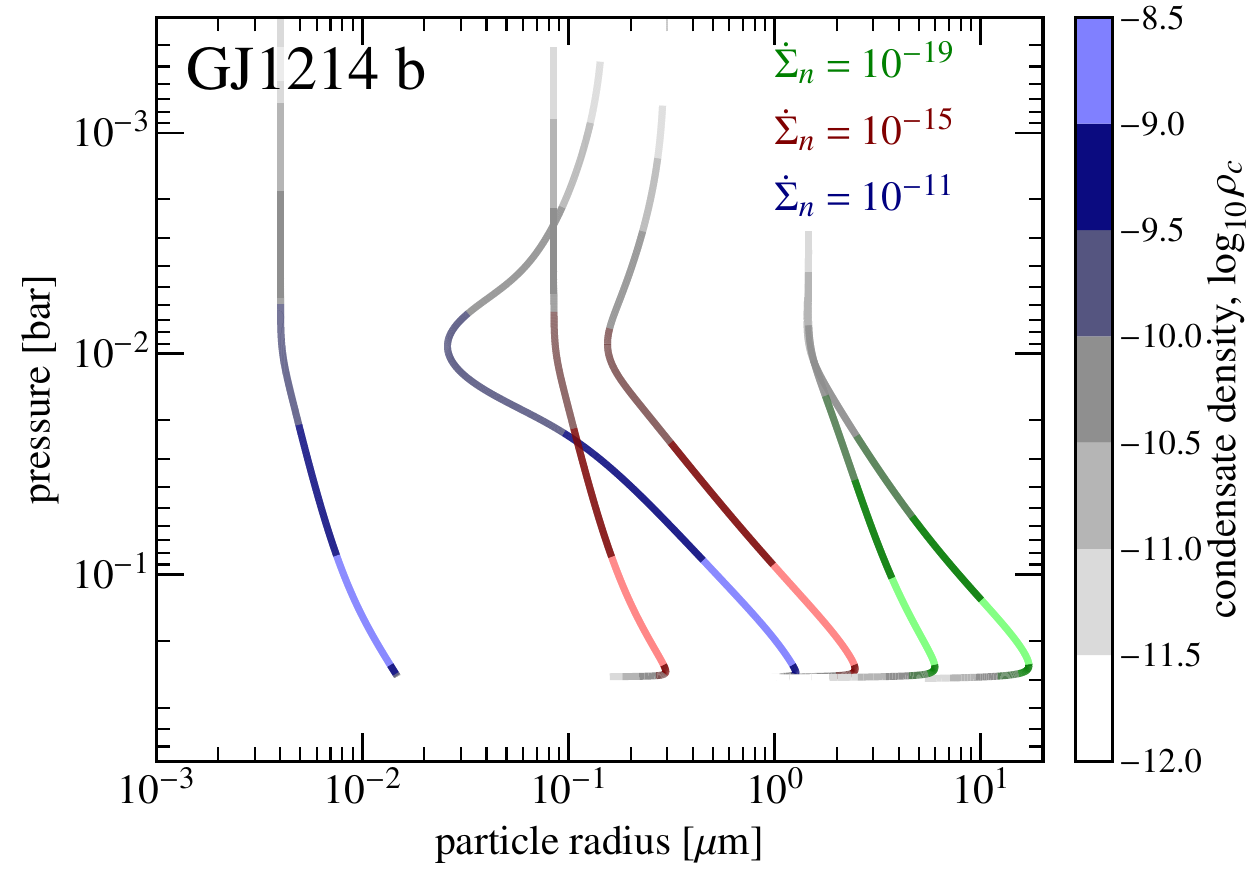}
    \caption{Obtained particle size as function of pressure for the cloud models applied to GJ1214b. The intensity of the rain in terms of the volume density of condensates ($\rho_c = x_c \rho_\mathrm{gas}$) is indicated by the color bar. Three values of the nuclei production rate are considered (as indicated by color) and results are plotted with and without accounting for coagulation. The diffusivity parameter is fixed at $K_\mathrm{zz}=10^8\,\mathrm{cm^2\, s^{-1}}$. Particle radii are larger in runs that include coagulation.}
    \label{fig:GJ1214b}
\end{figure}
GJ1214 b is a super-Earth or sub-Neptune planet of radius $R_p=2.7\pm0.1\,R_\oplus$ and mass $M_p=6.5\pm1.0\,M_\oplus$ orbiting an M4.5 star at a distance of 0.015\, au \citep{CharbonneauEtal2009}. With these bulk properties GJ1214 b could both a ``water world'' or a more standard terrestrial planet with a H/He envelope. Interestingly, GJ1214 b transmission spectra is virtually featureless \citep{KreidbergEtal2014}, indicative of clouds.

Cloud models have recently been applied to GJ1214 b \citep[e.g.,][]{GaoBenneke2018,OhnoOkuzumi2018}.  Here we apply our cloud model toward GJ1214 b with the aim of comparing the physical structure (particle sizes and concentrations) against these works in the broadest sense. A detailed comparison, let alone a calibration, is rather meaningless as these works employ vastly different cloud microphysical and atmospheric models.  

We consider KCl as our cloud species and use the $P_\mathrm{sat}$ profile presented in \citet{MorleyEtal2012}. The concentration of KCl at the bottom of the atmosphere is taken to be $x_{v,\mathrm{bot}}=3\times 10^{-4}$. We largely follow \citet{OhnoOkuzumi2018} in choosing our atmospheric parameters (see \tb{parameters}). However we keep $\kappa_\mathrm{IR}$ fixed; with $\kappa_\mathrm{IR}=0.03\,\mathrm{cm^2\,g^{-1}}$ we obtain a P-T profile (\fg{HJ-profile}) that resembles theirs.  The diffusivity is fixed at $K_\mathrm{zz}=10^{8}\,\mathrm{cm^2\,s^{-1}}$ while we consider the same three values for the nuclei production rate $\dot\Sigma_n$. Nuclei are injected at a height corresponding to a pressure of $0.01$ bar.

Results are shown in \fg{GJ1214b} where the particle radius is plotted against height for the three nucleation rates and for either the coagulation mode and the no-coagulation mode. The intensity of the rain in terms of the condensate volume density $\rho_c$ is indicated by the color. The rain reaches its highest intensity near the cloud base. Clearly, the nuclear production rate -- a free parameter in our model -- has a key influence on the grain size. Also, it can be seen that clouds with the smallest grains are also the most extended, since these grains tend to diffuse, rather than settle. Finally, grains are larger in runs where coagulation is included. These findings reflect the discussion of the hot generic hot-Jupiter clouds in \se{physical}.

Comparing these curves to the $K_\mathrm{zz}=10^{8}\,\mathrm{cm^2\,s^{-1}}$, 1x solar metallicity panel of Fig.~5 of \citet{GaoBenneke2018}, we see that their typical sizes of 1--10\,$\mu\mathrm{m}$ correspond well to our results with the low $\dot\Sigma_n$. (In their model the nucleation rate is given by a full microphysical model) The gradient in grain size with height seems to be a bit shallower in our models, however, considering that \citep{GaoBenneke2018} did not include coagulation. 

\citet{OhnoOkuzumi2018} also modeled GJ1214 b and, like us, used a characteristic size approach. In addition, they too prescribed the nucleation. However, they fixed the nuclei number density at the cloud base. Compared to our choice of prescribing the entire profile, this has the advantage of only introducing a single free parameter. On the other hand, it results in the largest grains residing in the top of the cloud, which seems somewhat spurious. Their typical grain size of 1--2 \micr nevertheless corresponds well to our results (they too account for coagulation) and their volume mixing ratios approach $x_c=10^{-4}$ -- the same as in our case.

\section{Transmission spectra (hot Jupiter)}

\begin{figure*}
    \centering
    \includegraphics[width=\textwidth]{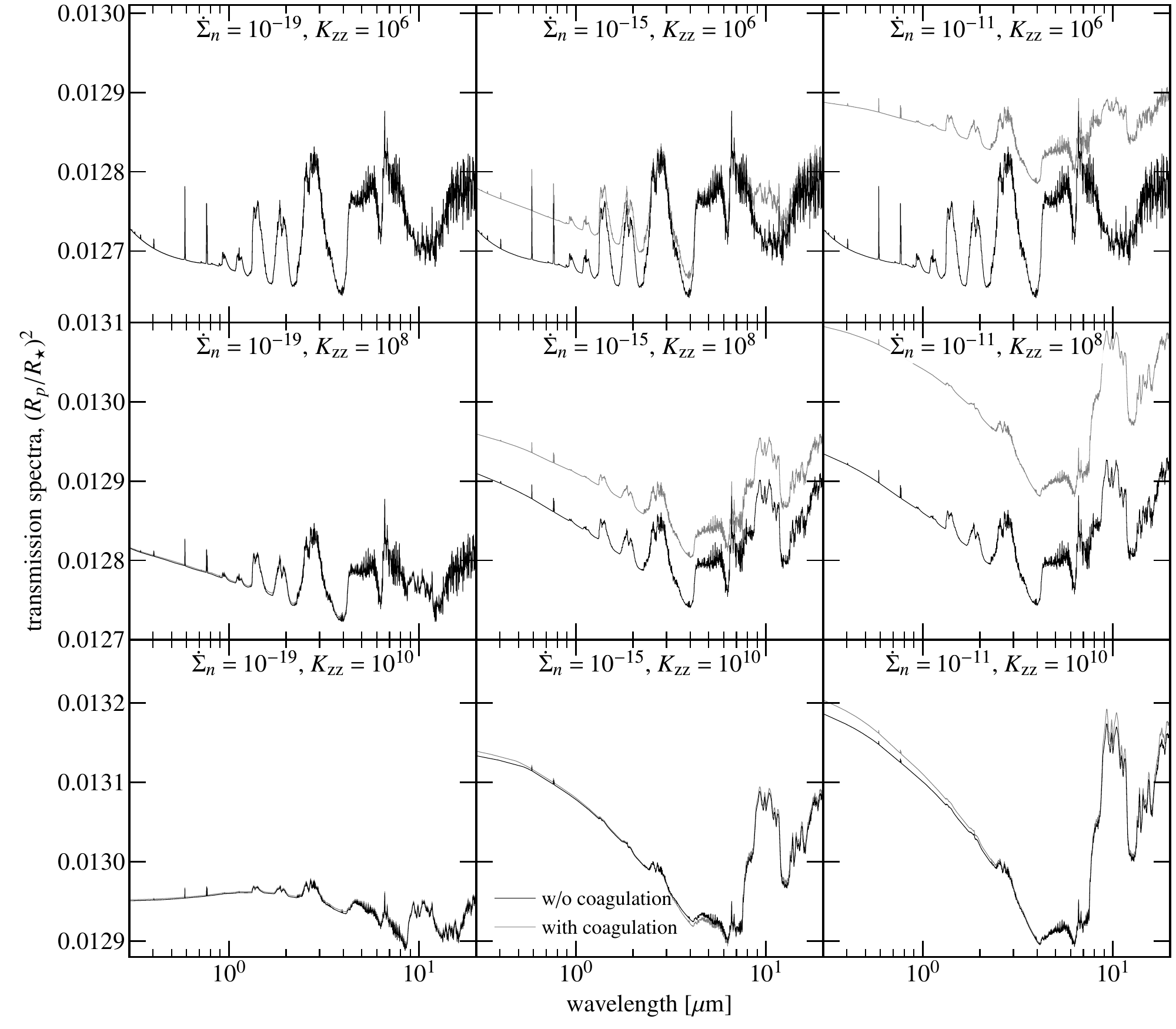}
    \caption{\texttt{ARCiS}-Simulated transmission spectra corresponding to the physical profiles presented in \fg{collage} as function of wavelength. Gray curves give the spectra without accounting for coagulation, black curves include coagulation. The scaling of the $y$-axis is different for the different diffusivities (higher $K_\mathrm{zz}$ results in a larger $R_p$).}
    \label{fig:collage-spectra}
\end{figure*}
\label{sec:spectra}
From \fg{collage} we see that the cloud thickness and particle size are heavily influenced by the diffusion strength and the nucleation rate. To investigate their effect on the spectral appearance of the transit signal of the planet, we computed for the nine cases of the generic hot-Jupiter model shown in \fg{collage} the transmission spectra. These are shown in \fg{collage-spectra}. 

To compute the spectra we have developed a radiative transfer tool for simulating exoplanet spectra. This code uses molecular line lists from the ExoMol project and the HITEMP and HITRAN databases to compute the molecular opacities with the method by \citet{2017A&A...607A...9M}. A validation of this \texttt{ARCiS} module is given in \app{validation}. Even though for the cloud condensation equations we use pure \mgsio as a condensate, we add 10\% of metallic iron to the particles when computing the optical properties. The implicit assumption is that the physical properties of the cloud particles (their sizes and concentrations) are well described by modeling the dominant condensate, in other words by our cloud model. However, this assumption cannot be made for the optical properties, which exhibit a strongly non-linear dependence on composition. \mgsio, for example,  is completely transparent in the near-IR, while only a small fraction of iron in the silicate lattice, or condensed inclusions, like metallic iron, will suffice to give a significant near-IR opacity. Hence, lacking a multi species cloud model, we account for this by adding a small amount of continuum opacity in the form of metallic iron. The 10\% metallic iron we take here is rather arbitrary and could be up to 50\% given the cosmic abundance of iron. The true iron fraction in planetary atmospheres is a parameter that we have to derive from observations or constrain from planet formation theory.

The cloud opacities are computed using refractive index data from \cite{1998A&A...339..904J} and \citet{1996A&A...311..291H} where we mix the iron and \mgsio together using effective medium theory. We use the distribution of hollow spheres (DHS) method from \citet{MinEtal2005} to convert the refractive index into particle optical properties. The gas phase chemistry is computed assuming thermochemical equilibrium using the code from \citet{MolliereEtal2017}. The atomic abundances that go into the chemical computations are assumed to be solar with depletions in Si, O, and Mg according to the computed value of $x_v$. This causes the C/O ratio to change in the cloud forming region, affecting the chemistry there. Below the cloud deck the C/O ratio is solar, $\mathrm{C/O}=0.55$, while in the cloud forming region $\mathrm{C/O}\approx0.7$.

In \fg{collage-spectra} the mid-IR transmission spectra are plotted for the same combination of diffusivities and nucleation rate as in \fg{collage}. The near- to mid-IR spectral region will become available with the MIRI instrument onboard JWST and further into the future with the recently selected ARIEL mission. Several inferences can be made. First, increasing the cloud thickness (either by increasing $\dot\Sigma_N$ or increasing $K_\mathrm{zz}$) suppresses the molecular features of, for example, H$_2$O in the 1--3 $\mu$m range. The reason is that, the $\tau=1$ height now resides much higher in the atmosphere to shield the molecular emission.

A striking result is the spectral appearance of \mgsio around 10 micron. The 10 micron silicate resonance is very sensitive to particle size. Small particles give a strong resonance signature, while increasing the particle size, the signature is flattened \citep[see e.g.,][]{MinEtal2005}. In addition, the solid feature stands out stronger against the (molecular) background for thicker clouds. Therefore, the resonance around 10 micron is most clearly seen in the case with high nucleation rate and diffusion strength (lower right panel), that is, a thick cloud of small particles. Only for the lowest diffusion strength (upper panels) does the silicate signature become unobservable around 10 micron. Finally, \fg{collage-spectra} displays a very interesting evolution of the slope of the near-IR signature. For the low nucleation rate models, the effect of increasing the cloud thickness (i.e., the diffusion strength) results in a gray near-IR spectrum. On the other hand, for the high nucleation rate, the near-IR spectrum is characterized by a much steeper slope. The reason behind this diverging trend with cloud thickness is the dependence of particle size with nucleation rate. Higher nucleation rates result in smaller grains whose opacity has a much steeper wavelength dependence in the near-IR region. Conversely, the 1--10 $\mu$m grains that are produced in the low $\dot{\Sigma}_n$, high $K$ run (bottom left panel) result in a gray opacity and a transmission spectra insensitive to wavelength.

The spectra we computed are sensitive to the effects of particle coagulation. The effects are twofold. One is that coagulation causes the grains to grow and settle deeper into the atmosphere. Second the opacity of the larger particles produced by coagulation is different. It can be seen that when we switch off the coagulation the cloud deck in the upper right four panels of \fg{collage-spectra} is much higher and thus mutes the molecular features more. In addition, the spectral appearance typical for small particles, the silicate feature at 10$\,\mu$m and the Rayleigh scattering slope at optical wavelengths, are reduced significantly by the effects of particle coagulation. While the case with low diffusion and high nucleation rate displays a strong cloud deck and silicate feature without coagulation, the spectral appearance is dominated by molecular features when coagulation is switched on. These considerations emphasize that  cloud features can only be properly interpreted by models that include coagulation.

\section{Model assessment}
\label{sec:assess}

We reflect on the achievement of our cloud model in the light of recent similar approaches. The key idea of our approach is to extend the simplicity and usability of the \citet{AckermanMarley2001} model with a more physical justified cloud model, while preserving its simplicity. The \citet{AckermanMarley2001} model already contained particle and vapor transport; however, it does not compute the size of the cloud particles. To proceed, a nucleation prescription is required. This we have done very crudely, simply by imposing it through ad-hoc prescriptions. Alternatively, nucleation can be treated from first principles. Photochemistry is a possible avenue for the formation of seed nuclei, which is thought to be the source of the haze as, for example, observed in Titan \citep{TomaskoEtal2005}. Another nucleation pathway is that of homogeneous nucleation, where the nuclei seed directly form out of the vapor. The hot interiors of exoplanets characterized by thick envelopes will guarantee evaporation of any condensate at some depth. For these planets homogeneous condensation may be considered the natural way to form clouds.

These additions to the nucleation model can, in principle, render the model more physically rigorous. However they also come at a drawback.  A well-known issue with the classical nucleation theory is that it mispredicts nucleation rates by many orders of magnitude \citep[e.g.,][]{FederEtal1966,TanakaEtal2005i,HorschEtal2008,DiemandEtal2013}. 
Similarly, codes that model haze formation necessarily rely on a large chemical network, with hundreds of reactions, and sophisticated radiation transport \citep{LavvasEtal2008,LavvasKoskinen2017,KawashimaIkoma2018}.  
Obviously, parametrizing nucleation implies that the size of the typical cloud particle no longer follows from first principles. But the transport model still addresses variations of particle concentration and size with height, which act as independent model constraints.

Another major simplification we have adopted is the characteristic particle approach (as in \citealt{OhnoOkuzumi2017}). A brief discussion on the validity and limitations of this approach can be found in \citet{KawashimaIkoma2018}. Recently, several studies have used CARMA\footnote{Community Aerosol and Radiation Model for Atmospheres.} toward modeling clouds on exoplanets \citep{GaoEtal2018,GaoBenneke2018,PowellEtal2018}. An output of this code is the particle size distribution at any height. A possible approach is to reconstruct the entire grain size distribution from the characteristic size (cf.\ \citealt{2008A&A...485..547H} or \citealt{BirnstielEtal2012} for disks). Nevertheless, within the \texttt{ARCiS} framework, solving for the particle size distribution is too computationally intensive, since we intend it to be used in future MCMC parameter searches. 
Altogether, we make no claim to have invented the ``best'' cloud model in terms of physical rigor, but one that is minimalistic, physical consistent and above all useful. Its modular approach can easily be extended to include more physical processes and its results can guide sophisticated, computationally intensive models in a complementary fashion.

\section{Summary}
\label{sec:summary}
In this paper we have studied the effects of diffusion strength and nucleation efficiency on the characteristics of clouds in exoplanet atmospheres. We have presented a relatively simple framework of cloud formation where these effects can be studied efficiently. Both the nucleation rate and the diffusion strength are key parameters in determining the properties of the cloud particles and the extent of the cloud. Since both these parameters are highly uncertain, it is important to understand their effects. We have presented simulated infrared transmission spectra for different combinations of these two parameters in a typical hot Jupiter atmosphere.

\noindent For the physical structure of the clouds we conclude that:
\begin{itemize}
	\item Increasing the nucleation rate results in thicker clouds of smaller particles. The high number of nuclei facilitate condensation. At the same time, the condensed mass is distributed over a larger number of particles, resulting in on average smaller particles.
    \item Increasing the diffusion strength results in thicker clouds. In this case, more vapor is mixed up and can condense on the nuclei. This causes simply more cloud material at each altitude and thus thicker clouds.
\end{itemize}

\noindent For the transmission spectra resulting from these structures we conclude:
\begin{itemize}
    \item For increasing diffusion strength and to a lesser degree increasing nucleation rate the molecular features weaken. This is caused by increasingly thicker clouds shielding more of the gaseous atmosphere.
    \item For high values of the diffusion strength and nucleation rate, the solid state 10 $\mu$m silicate feature appears. This feature of the cloud particles is visible in almost all parameter settings we consider here, but is most prominent for the highest values of diffusion and nucleation because they create the thickest clouds with small particles.
    \item For increasing nucleation rates, that is, smaller particles, the slope of the Near-IR steepens.
    \item Coagulation has a significant influence on the spectral appearance of the clouds, especially in the case of high nucleation rates.
\end{itemize}

The above observational features can be used to characterize cloud particles in exoplanet atmospheres. The modeling framework we present in this paper is computationally not very demanding. We can see two very important extensions of the present model. First, the opacities obtained from the cloud model can be fed back to the physical structure, such that for example the temperature profile is obtained self-consistently (recall that we used a fixed $\kappa_\mathrm{IR}$ in calculating the $P$-$T$ profile). Second, and maybe more important, we can use this modeling framework to include a physically motivated cloud formation model in retrieval methods. This way we can simulate the effect that clouds have on the atmospheric composition and observational features. In addition, it allows us to put observational constraints on physical parameters like the nucleation rate and the diffusion strength. This will provide a significant step forward in understanding the physical processes in exoplanet atmospheres.

\begin{acknowledgements}
C.W.O.\ and M.M.\ are grateful to Paul Molli\`ere for providing his \texttt{petitCODE} to calculate the gas composition. The authors also acknowledge fruitful discussion with Christiane Helling, Paul Molli\`ere, and Peter Woitke. C.W.O.\ is supported by the Netherlands Organization for Scientific Research (NWO; VIDI project 639.042.422).  The research leading to these results has received funding from the European Union’s Horizon 2020 Research and Innovation Programme, under Grant Agreement 776403.
\end{acknowledgements}

\bibliographystyle{aa}
\bibliography{rain}

\appendix
%\begin{appendix}

\section{Validation with \texttt{petitCODE}}
\label{sec:validation}

We have validated the computations performed with \texttt{ARCiS} with the exoplanet simulation code \texttt{petitCODE} \citep{MolliereEtal2015, MolliereEtal2017}. The \texttt{petitCODE} has been extensively benchmarked in \cite{BaudinoEtal2017}. We compute the transmission spectrum of the atmospheric setup from the model used here without any cloud formation. The chemistry, hydrostatic structure, molecular opacities and resulting transmission spectrum are computed both by \texttt{ARCiS} and \texttt{petitCODE} independently. The chemical equilibrium module used in \texttt{ARCiS} is the same as the one used in \texttt{petitCODE}. This module is benchmarked in \cite{BaudinoEtal2017}, so here we only check the proper implementation of the module in \texttt{ARCiS}. \Fg{petitCODE_ARCiS} shows the comparison of the resulting transmission spectra. The spectra match exceptionally well at almost all wavelengths. There are some small differences in the optical part of the spectrum which can be attributed to a different Rayleigh scattering law and different opacities for TiO and VO used in both codes. These differences are irrelevant for the purpose of this paper. The \texttt{petitCODE} spectrum is computed at higher spectral resolution and thus shows small high frequency variations which are smoothed in the lower resolution \texttt{ARCiS} spectrum. We conclude that the match is excellent.

\begin{figure}
    \centering
    \includegraphics[width=\columnwidth]{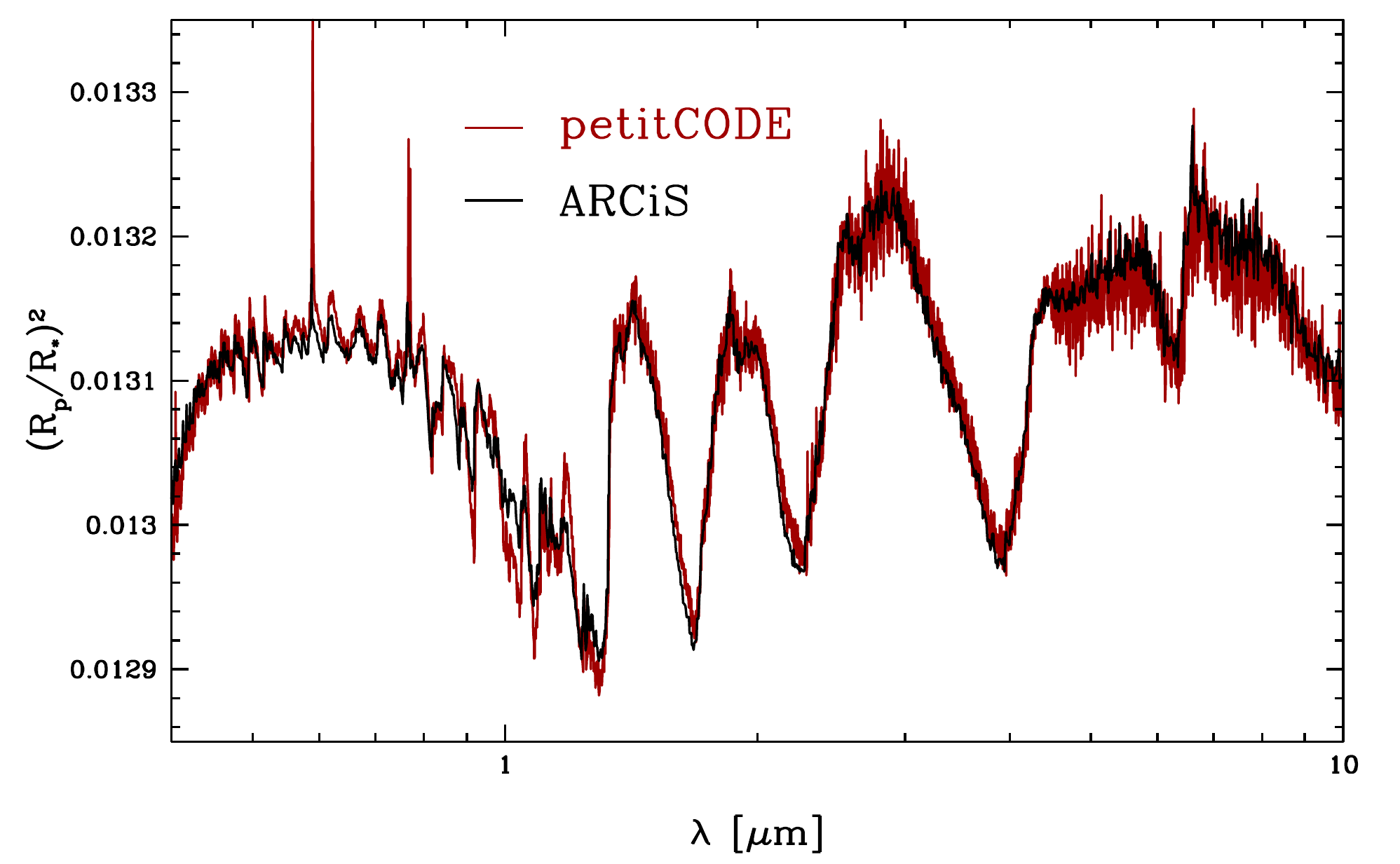}
    \caption{Transmission spectra for the standard model without clouds computed using \texttt{petitCODE} and \texttt{ARCiS}.}
    \label{fig:petitCODE_ARCiS}
\end{figure}
%\end{appendix}

\end{document}